\documentstyle[pre,aps]{revtex}

   \font\tenmsb=msbm10 scaled\magstep 1
   \font\sevenmsb=msbm7 scaled \magstep 1
   \font\faivemsb=msbm5 scaled \magstep 1
\newfam\msbfam
      \textfont\msbfam=\tenmsb
      \scriptfont\msbfam=\sevenmsb
      \scriptscriptfont\msbfam=\faivemsb
\def\Bbb#1{{\fam\msbfam #1}}
\font\tengothic=eufm10 scaled\magstep 1
\font\sevengothic=eufm7 scaled\magstep 1
\newfam\gothicfam
      \textfont\gothicfam=\tengothic
      \scriptfont\gothicfam=\sevengothic

\newcommand{\be}{\begin{equation}}
\newcommand{\ee}{\end{equation}}
\newcommand{\sgm}{\sigma}
\newcommand{\dlt}{\delta}
\newcommand{\ra}{\rightarrow}
\newcommand{\lbd}{\lambda}
\newcommand{\bt}{\beta}
\newcommand{\prt}{\partial}
\newcommand{\vp}{\varphi}
\newcommand{\al}{\alpha}
\newcommand{\gm}{\gamma}
\newcommand{\om}{\omega}
\newcommand{\br}{{\bf r}}

\begin{document}

\draft

\title{Stochastic Instability of Quasi-Isolated Systems}

\author{V.I. Yukalov}

\address{Research Center for Optics and Photonics \\
Instituto de Fisica de S\~ao Carlos, Universidade de S\~ao Paulo \\
Caixa Postal 369, S\~ao Carlos, S\~ao Paulo 13560-970, Brazil \\
and \\
Bogolubov Laboratory of Theoretical Physics \\
Joint Institute for Nuclear Research, Dubna 141980, Russia}

\maketitle

\begin{abstract}

The stability of solutions to evolution equations with respect to 
small stochastic perturbations is considered. The stability of a 
stochastic dynamical system is characterized by the local stability 
index. The limit of this index with respect to infinite time describes
the asymptotic stability of a stochastic dynamical system. Another 
limit of the stability index is given by the vanishing intensity of
stochastic perturbations. A dynamical system is stochastically unstable
when these two limits do not commute with each other. Several examples
illustrate the thesis that there always exist such stochastic 
perturbations which render a given dynamical system stochastically
unstable. The stochastic instability of quasi-isolated systems is
responsible for the irreversibility of time arrow.

\end{abstract}

\vskip 1cm


\section{Introduction}

Evolutional processes of nature are described by differential equations
that, in general, are equations in partial derivatives. A set of such 
partial differential equations constitutes an infinite-dimensional
{\it dynamical system}. Under a {\it physical system} one implies en
ensemble of objects whose behaviour is of interest. The evolution of a 
given physical system is characterized by the related dynamical system.
Among physical systems, one distinguishes {\it isolated systems} as
opposed to {\it open systems}. The evolution of the isolated physical 
systems is governed by deterministic laws, that is, by {\it deterministic 
equations}, not containing random variables. While open physical systems,
generally, deal with {\it stochastic equations}, where random terms 
represent the interaction with surrounding.

Solutions to differential equations can be either stable or unstable.
There are methods for analyzing the stability of solutions for a 
given dynamical system, either deterministic [1--3] or stochastic [4]. 
Here we address another problem, that of stability of a deterministic 
dynamical system with respect to small stochastic perturbations. This 
problem is not only interesting by itself but it is of fundamental 
importance with regard to the question: How adequately the notion of 
isolated systems represents the physical reality?

As is evident, the notion of an isolated system is an abstraction. In 
fact, no real system can be completely isolated from its surrounding. 
This point has been repeatedly emphasized in literature [5--9]. And 
the impossibility of ideally isolating macroscopic systems from their 
environment is considered as being intimately related with the
irreversibility of time [10,11]. Moreover, it has been stressed [12,13] 
that the concept of an isolated system is logically self-contradictory 
by its own. This is because to realize the isolation, one has to employ 
technical devices acting on the system; and to ensure that the latter 
is kept isolated, one must apply measuring instruments perturbing the 
system. The preparation and registration processes disturb the system 
dynamics [14]. In this way, there exists an accepted understanding that 
any considered physical system is never absolutely isolated but is 
subject to, probably, weak but, generally, uncontrollable random 
influence from the environment. Even if this influence is quite weak, 
its very existence is of principal importance, for explaining the 
irreversibility of time arrow.

It is worth noting that the irreversible behaviour of macroscopic 
systems is often attributed to internal chaotic nature of microscopic
dynamics (see discussion in [15]). However, not all physical systems
display chaotic behaviour. Many of them are perfectly governed by
rather simple deterministic laws, with no signs of chaos. Nevertheless,
the time arrow is well defined for any system, including very simple 
and not chaotic ones. What is more, the recent developments in dynamical
theory, as reviewed by Zaslavsky [16], show that chaotic dynamics in real
systems does not provide finite relaxation time to equilibrium or fast
decay of fluctuations, and that chaotic systems are not completely 
random in the sense originally postulated for statistical systems. 
Therefore the presence of random environment, though very weak, seems 
to be crucially important for interpreting fundamental notions in the
behaviour of real physical systems.

From another side, there is a common belief, based on practical 
experience, that physical systems can, with a very good accuracy, be
isolated and can be described by deterministic equations, while the
random influence of surrounding may be neglected. Thus, there exists 
an apparent contradiction  between the principal necessity of allowing 
for random perturbations influencing any real system and the practical
possibility of neglecting such perturbations, treating a system as
isolated.

This contradiction is resolved in the present paper by putting the 
problem on a firm mathematical footing. The concept of quasi-isolated
systems is defined. It is shown that such systems, generally, are
unstable with respect to infinitesimally small stochastic perturbations.
At the same time, for a finite temporal period, these systems can be 
treated as approximately isolated.

\section{Stability of Stochastic Systems}

Let a continuous variable $x\in\Bbb{D}$ denote a set of spatial 
coordinates pertaining to a domain $\Bbb{D}$ and let $t\in\Bbb{R}_+$
denote time. Suppose a stochastic field $\xi(t)$ is defined. In 
general, the latter is a set of stochastic functions $\xi_i(x,t)$,
with $i=1,2,\ldots$. Throughout the paper, we shall use the matrix
notation [17] making it possible to express the following equations in 
a compact form. Thus, the {\it stochastic field} $\xi(t)=[\xi_i(x,t)]$
is considered as  a column with respect to both $i=1,2,\ldots$ as well
as $x\in\Bbb{D}$. The {\it dynamical state} $y(\xi,t)=[y_i(x,\xi,t)]$ is
also a column with respect to $i$ and $x$, as is the {\it velocity
field} $v(y,\xi,t)=[v_i(x,y,\xi,t)]$. The set of evolution equations,
defining a dynamical system, in the matrix notation reads
\be
\label{1}
\frac{d}{dt}\; y(\xi,t) = v(y,\xi,t) \; .
\ee
This is complimented by an initial condition
\be
\label{2}
y(\xi,0) = y(0) \; ,
\ee
implying the set
$$
y_i(x,\xi,0) = y_i(x,0) \qquad (i=1,2,\ldots)
$$
of the related initial conditions. The averaging over the stochastic 
field $\xi(t)$ is denoted by the double angle brackets as
\be
\label{3}
y(t) = \ll y(\xi,t) \gg \; ,
\ee
which assumes the family of the functions
\be
\label{4}
y_i(x,t) =\; \ll y_i(x,\xi,t) \gg \; ,
\ee
with $i=1,2,\ldots$.

In the stochastic equation (1), the velocity field $v(y,\xi,t)$ 
may, in general, contain differential as well as integral operations. 
To solve Eq. (1) means to find the averaged solution (3). Stochastic
differential equations, as is known [18], can be defined either in 
the sense of Ito or in the sense of Stratonovich. In what follows, 
the latter definition will be employed, which permits simpler 
calculations and is better motivated physically [19]. It is also
possible to use the {\it stochastic expansion technique} [20,21],
presenting the stochastic field as an expansion over smooth functions 
of spatial and temporal variables with random coefficients. This
method enables the usage of the standard differential and integration 
analysis. The final results of the expansion technique coincide with
the corresponding expressions obtained by means of the Stratonovich
method.

The local stability of a dynamical system can be characterized by the
{\it local stability index}
\be
\label{5}
\sgm(t) \equiv \ln\sup_{\dlt y(0)}\; 
\frac{|\dlt y(t)|}{|\dlt y(0)|} \; ,
\ee
which describes the maximal deviation of the averaged trajectory at
time $t$ after an infinitesimal variation of the initial conditions.
Such a deviation, according to Eq. (5), corresponds to the law
\be
\label{6}
|\dlt y(t)| \sim |\dlt y(0)| e^{\sgm(t)} \; ,
\ee
from where it is evident why $\sgm(t)$ is called the stability index,
or stability exponent. From this definition, one can immediately 
conclude that the admissible local properties of motion are classified
as:
$$
\sgm(t) < 0 \qquad (locally\; stable) \; ,
$$
$$
\sgm(t) = 0 \qquad (locally \; neutral) \; ,
$$
\be
\label{7}
\sgm(t) > 0 \qquad (locally \; unstable) \; .
\ee
The asymptotic Lyapunov stability corresponds to the terminology:
$$
\lim_{t\ra\infty} \sgm(t) = -\infty \qquad (Lyapunov\; stable)\; ,
$$
\be
\label{8}
\lim_{t\ra\infty} \sgm(t) > -\infty \qquad (Lyapunov\; unstable)\; .
\ee
And in the language of the Lagrange stability of motion, one has:
$$
\sup_{t} \sgm(t) < \infty \qquad (Lagrange \; stable)\; ,
$$
\be
\label{9}
\sup_{t} \sgm(t) = \infty \qquad (Lagrange\; unstable)\; .
\ee
The limit
\be
\label{10}
\lbd =\lim_{t\ra\infty} \; \frac{1}{t}\;\sgm(t)
\ee
corresponds to the largest Lyapunov exponent. One tells that the 
motion is asymptotically stable if $\lbd<0$, neutral when $\lbd=0$, 
and unstable if $\lbd>0$.

The usage of a local characteristic of motion, such as the local 
stability index (5), provides us an essentially richer information
on temporal dynamics than the largest Lyapunov exponent (10) defined
for the limit $t\ra\infty$. First of all, this is because many 
dynamical systems possess a rather complicated structure of their 
phase space resembling a topological zoo, consisting of domains of 
chaotic dynamics as well as of regions of regular motion, containing
manifolds of wandering trajectories as well as trapping islands. As
a result of this, the fine local properties of orbits play a leading 
role, while such a fairly rough characteristic as the limiting Lyapunov
exponent is less important [16,22].

Moreover, the asymptotic divergence of trajectories of stochastic 
dynamical systems is not compulsory exponential [4], because of which
making use of only the limiting Lyapunov exponent (10) may result in 
the loss of information. For example, the divergence of trajectories 
can be of power law
$$
|\dlt y(t)| \sim |\dlt y(0)| t^\bt \; .
$$
Such power laws are typical for weakly disordered systems [23] 
exhibiting mid-range order [24]. In that case, the local stability 
index (5) behaves as $\sgm(t)\sim\bt\ln t$, which can be either 
positive or negative depending on the sign of $\bt$. Respectively, 
the motion is either stable or unstable. While, according to the 
Lyapunov exponent (10), which is $\lbd=0$, the motion is neutral.
Another example has to do with the divergence of trajectories by the
stretched exponential law
$$
|\dlt y(t)| \sim |\dlt y(0)|\exp\left ( \kappa t^\bt\right ) \; ,
$$
with $0<\bt<1$, which is also quite ubiquitous in disordered systems.
Then the local stability index (5) is $\sgm(t)\sim\kappa t^\bt$, which
again can be either positive or negative depending on the sign of
$\kappa$, hence, the motion is either stable or unstable. And the limit
(10) is again zero, classifying the motion as neutral.

Instead of the asymptotic Lyapunov exponent (10), one could define the
local Lyapunov exponent [25,26] as
$$
\lbd (t) = \frac{1}{t}\; \sgm(t) \; .
$$
However, for what follows, the usage of the local stability index (5)
is more convenient.

One more advantage of employing a local characteristic of stability 
is that the limit (10) for many complex systems is technically 
unachievable. Then the local index (5) is the sole available quantity 
that can be actually calculated. Such a situation is typical for
complicated nonlinear equations that can be treated only numerically
[27], for the analysis of various time series that are always finite
[28], and for the dynamical representation of perturbation theory,
where it is practically feasible to calculate only a finite number of
terms [29--32].

The local stability exponent (5) can be expressed through the multiplier
matrix $\hat M(t)=[M_{ij}(x,x',t)]$ with the elements
\be
\label{11}
M_{ij}(x,x',t) \equiv \frac{\dlt y_i(x,t)}{\dlt y_j(x',0)}\; .
\ee
From this definition, it follows that
\be
\label{12}
M_{ij}(x,x',0) = \dlt_{ij}\dlt(x-x') \; ,
\ee
where $\dlt_{ij}$ is the Kroneker delta and $\dlt(x)$ is the Dirac
delta-function. Writing the variation of the averaged dynamic state as
\be
\label{13}
\dlt y(t)  =\hat M(t)\dlt y(0) \; ,
\ee
we see that
\be
\label{14}
\sup_{\dlt y(0)}\; \frac{|\hat M(t)\dlt y(0)|}{|\dlt y(0)|} =
||\hat M(t)|| \; ,
\ee
with the spectral norm of $\hat M(t)$ being assumed. Therefore the
local stability exponent (5) is
\be
\label{15}
\sgm(t) = \ln||\hat M(t)|| \; .
\ee
Thus, to analyse the stability of motion, we need to know the 
multiplier matrix (11).

\section{Stochastic Multiplier Matrix}

What we are actually given is the stochastic equation (1) defining 
the stochastic dynamic state $y(\xi,t)$, whose variation
\be
\label{16}
\dlt y(\xi,t) = \hat M(\xi,t)\dlt y(0) 
\ee
over the initial conditions involves the {\it stochastic multiplier 
matrix} $\hat M(\xi,t)=[M_{ij}(x,x',\xi,t)]$ with the elements
\be
\label{17}
M_{ij}(x,x',\xi,t) \equiv 
\frac{\dlt y_i(x,\xi,t)}{\dlt y_j(x',0)} \; .
\ee
For the latter, one has the initial condition
\be
\label{18}
M_{ij}(x,x',\xi,0) = \dlt_{ij}\; \dlt(x-x') \; .
\ee

The multiplier matrix (17) is connected with the {\it stochastic 
Jacobian matrix} $\hat J(\xi,t)=[J_{ij}(x,x',\xi,t)]$ with the elements
\be
\label{19}
J_{ij}(x,x',\xi,t) \equiv 
\frac{\dlt v_i(x,y,\xi,t)}{\dlt y_j(x',\xi,t)} \; .
\ee
The variational differentiation of Eq. (1) gives the equation
\be
\label{20}
\frac{d}{dt}\; \hat M(\xi,t) = \hat J(\xi,t)\; \hat M(\xi,t)
\ee
for the multiplier matrix (17). The initial condition for this 
equation is Eq. (18).

Since the evolution equation (1) represents a set of partial 
differential equations, one has to define as well boundary conditions. 
The latter can be written in the general form
\be
\label{21}
b(y,\xi,t) = 0 \qquad (x\in \prt\Bbb{D}) \; ,
\ee
where $\prt\Bbb{D}$ is the boundary manifold of the domain $\Bbb{D}$ 
and $b(y,\xi,t)=[b_i(x,y,\xi,t)]$ is a boundary vector. Defining
the {\it boundary matrix} $\hat B(\xi,t)=[B_{ij}(x,x',\xi,t)]$ with 
the elements
\be
\label{22}
B_{ij}(x,x',\xi,t) \equiv 
\frac{\dlt b_i(x,y,\xi,t)}{\dlt y_j(x',\xi,t)}
\ee
and accomplishing the variation of Eq. (21), we get the boundary
condition
\be
\label{23}
\hat B(\xi,t)\hat M(\xi,t) = 0 \qquad (x\in\prt\Bbb{D})
\ee
for the multiplier matrix.

As an illustration, we may offer the often met form of the boundary
conditions
$$
\left ( 1 + \zeta\; \frac{\prt}{\prt x}\right ) y_i(x,\xi,t) =
f_i(t) \qquad (x\in\prt\Bbb{D}) \; ,
$$
where $\zeta$ is a parameter and $f_i(t)$ is a given function. The
variation of this condition results in the equation
$$
\left ( 1 + \zeta\;\frac{\prt}{\prt x}\right )
M_{ij}(x,x',\xi,t) = 0 \qquad (x\in\prt\Bbb{D}) \; ,
$$
demonstrating a particular case of the boundary condition (23).

For the multiplier and Jacobian matrices, one may employ different 
representations. To this end, let a set $\{\vp_n(t)\}$ of the columns
$\vp_n(t)=[\vp_{ni}(x,t)]$ be given, forming an orthonormalized
complete basis,
$$
\vp^+_m(t)\; \vp_n(t) =\dlt_{mn} \; , \qquad
\sum_n \vp_n(t)\; \vp_n^+(t) =\hat 1 \; ,
$$
where $\hat 1=[\dlt_{ij}\dlt(x-x')]$ is the unity matrix and $n$ is
a labelling multi-index. To pass from the $x$-representation to
$n$-representation, we define
\be
\label{24}
M_{mn}(\xi,t) \equiv \vp_m^+(t)\hat M(\xi,t)\vp_n(t) \; , \qquad
J_{mn}(\xi,t) \equiv \vp_m^+(t)\hat J(\xi,t)\vp_n(t) \; .
\ee
Recall that the matrix notation [17] is used here, according to which,
for instance, the action of the multiplier matrix on $\vp_n(t)$ is the
column
$$
\hat M(\xi,t)\vp_n(t) =\left [ \sum_j \int
M_{ij}(x,x',\xi,t) \vp_{nj}(x',t)\; dx'\right ] \; .
$$
Equation (20) for the multiplier matrix in the new representation reads
\be
\label{25}
\frac{d}{dt}\; M_{mn}(\xi,t) = \sum_k \left [ 
J_{mk}(\xi,t)M_{kn}(\xi,t) + M_{mk}(\xi,t) \vp_k^+(t)\;
\frac{d\vp_n(t)}{dt} - \vp_m^+(t)\; \frac{d\vp_k(t)}{dt}\;
M_{kn}(\xi,t)\right ] \; ,
\ee
where the relation
$$
\frac{d\vp_m^+(t)}{dt} \; \vp_n(t) + \vp_m^+(t)\;
\frac{d\vp_n(t)}{dt} = 0 \; ,
$$
following from the normalization condition, is used. And from Eq. (18),
we have the initial  condition
\be
\label{26}
M_{mn}(\xi,0) =\dlt_{mn} 
\ee
for Eq. (25). The multiplier matrix enjoys several useful properties.

\vskip 2mm

{\bf Proposition 1}. If the dynamical state $y(\xi,t)$ can be presented 
as an expansion
\be
\label{27}
y(\xi,t) = \sum_n c_n(\xi,t)\; \vp_n(t) + f(t) \; ,
\ee
over a basis $\{\vp_n(t)\}$ and $f(t)=[f_i(x,t)]$ is a column 
of functions not depending on the initial state $y(0)$, then the 
multiplier matrix has the form
\be
\label{28}
\hat M(\xi,t) = \sum_n \mu_n(\xi,t)\; \vp_n(t)\; \vp_n^+(0)\; ,
\ee
in which
\be
\label{29}
\mu_n(\xi,t) \equiv \frac{\dlt c_n(\xi,t)}{\dlt c_n(\xi,0)} \; .
\ee

\vskip 2mm

{\bf Proof}. The variation of the expansion (27) gives
$$
\frac{\dlt y_i(x,\xi,t)}{\dlt y_j(x',0)} = \sum_n
\frac{\dlt c_n(\xi,t)}{\dlt c_n(\xi,0)}\;
\frac{\dlt c_n(\xi,0)}{\dlt y_j(x',0)}\; \vp_{ni}(x,t) \; .
$$
At the same time, from Eq. (27) we have
$$
c_n(\xi,t) = \vp_n^+(t) y(\xi,t) - \vp_n^+(t)f(t) \; .
$$
From the latter equation, we get
$$
\frac{\dlt c_n(\xi,0)}{\dlt y_j(x',0)} = \vp_{nj}^*(x',0) \; .
$$
Using this and invoking the definition (17), we obtain the form (28) 
with notation (29).

\vskip 2mm

{\bf Remarks}. Although the basis $\{\vp_n(t)\}$ is assumed to be 
orthonormalized, but the vectors $\vp_m(t_1)$ and $\vp_n(t_2)$ at
different times $t_1\neq t_2$ are not necessarily orthogonal, so that,
in general,
$$
\vp_m^+(0)\; \vp_n(t)\neq \dlt_{mn} \; .
$$
Neither $\vp_n(t)$ nor $\vp_n(0)$ are necessarily the eigenvectors of
the multiplier matrix, for which we have
$$
\hat M(\xi,t)\; \vp_n(0) = \mu_n(\xi,t)\vp_n(t) \; .
$$
Only when $\vp_n(t)=\vp_n$ does not depend on time, then $\vp_n$ is
an eigenvector of $\hat M(\xi,t)$ and $\mu_n(\xi,t)$ is its eigenvalue.

\vskip 2mm

{\bf Proposition 2}. Suppose the multiplier matrix $\hat M(\xi,t)$ 
possesses eigenvectors $\vp_n(t)$ forming a complete orthonormalized
basis. Then the related eigenvalues, given by the eigenproblem
\be
\label{30}
\hat M(\xi,t)\vp_n(t) = \mu_n(\xi,t)\vp_n(t) \; ,
\ee
can be presented as
\be
\label{31}
\mu_n(\xi,t) =\exp\left\{ \int_0^t J_{nn}(\xi,t')\; dt'\right\} \; .
\ee

\vskip 2mm

{\bf Proof}. With $\vp_n(t)$ being the eigenvectors of the multiplier
matrix, the elements of the latter, defined in Eq. (24), are
\be
\label{32}
M_{mn}(\xi,t) = \dlt_{mn}\mu_n(\xi,t) \; .
\ee
Substituting this into Eq. (25) yields
\be
\label{33}
\dlt_{mn}\; \frac{d}{dt}\; \mu_n(\xi,t) =  J_{mn}(\xi,t)\mu_n(\xi,t) +
[ \mu_m(\xi,t) - \mu_n(\xi,t)]\vp_m^+(t)\; \frac{d\vp_n(t)}{dt} \; .
\ee
When $m=n$, the latter equation gives
\be
\label{34}
\frac{d}{dt}\; \mu_n(\xi,t) = J_{nn}(\xi,t)\mu_n(\xi,t) \; ,
\ee
while for $m\neq n$, it results in
$$
J_{mn}(\xi,t) = \left [ 1 -\; \frac{\mu_m(\xi,t)}{\mu_n(\xi,t)}
\right ] \vp_m^+(t) \; \frac{d\vp_n(t)}{dt} \; .
$$
Solving Eq. (34), with the initial condition
\be
\label{35}
\mu_n(\xi,0) = 1 \; ,
\ee
we come to the eigenvalue (31).

\vskip 2mm

{\bf Remarks}. From the eigenproblem (30), one gets the representation
\be
\label{36}
\hat M(\xi,t) =\sum_n \mu_n(\xi,t)\vp_n(t)\vp_n^+(t) 
\ee
for the multiplier matrix. The eigenvectors of the latter are not 
necessarily the eigenvectors of the Jacobian matrix (19). Hence the
form $J_{mn}(\xi,t)$, defined in Eq. (24), is, in general, nondiagonal.

\vskip 2mm

{\bf Proposition 3}. Assume that a complete orthonormalized basis 
$\{\vp_n(t)\}$ is such that
\be
\label{37}
\vp_m^+(t)\; \frac{d\vp_n(t)}{dt} = 0 \qquad (m\neq n) \; .
\ee
Then $\vp_n(t)$ are the eigenvectors of the multiplier matrix 
$\hat M(\xi,t)$ if and only if they are also the eigenvectors of
the Jacobian matrix $\hat J(\xi,t)$.

\vskip 2mm

{\bf Proof}. Let condition (37) hold. Then Eq. (25) becomes
\be
\label{38}
\frac{d}{dt}\; M_{mn}(\xi,t) = \sum_k J_{mk}(\xi,t) M_{kn}(\xi,t) +
M_{mn}(\xi,t) \left [ \vp_n^+(t)\; \frac{d\vp_n(t)}{dt} \; - 
\vp_m^+(t)\; \frac{d\vp_m(t)}{dt}\right ] \; .
\ee
If $\vp_n(t)$ are the eigenvectors of $\hat M(\xi,t)$, that is, the
form (32) takes place, then Eq. (38) reduces to 
$$
\dlt_{mn}\; \frac{d}{dt}\; \mu_n(\xi,t) = 
J_{mn}(\xi,t)\mu_n(\xi,t) \; ,
$$
from where it is clear that
\be
\label{39}
J_{mn}(\xi,t) = \dlt_{mn} J_{nn}(\xi,t) \; .
\ee
Hence, $\vp_n(t)$ are the eigenvectors of $\hat J(\xi,t)$.

Conversely, if $\vp_n(t)$ are the eigenvectors of $\hat J(\xi,t)$,
so that Eq. (39) holds true, then solving Eq. (38) yields
$$
M_{mn}(\xi,t) = M_{mn}(\xi,0)\exp\left\{ \int_0^t \left [
J_{mm}(\xi,t') + \vp_n^+(t')\; \frac{d\vp_n(t')}{dt'} \; -
\vp_m^+(t')\; \frac{d\vp_m(t')}{dt'}\right ] dt'\right \} \; .
$$
In view of the initial condition (26), this results in
\be
\label{40}
M_{mn}(\xi,t) = \dlt_{mn}\exp \left\{ \int_0^t J_{nn}(\xi,t')\; dt'
\right \} \; ,
\ee
which tells us that $\vp_n(t)$ are the eigenvectors of $\hat M(\xi,t)$.

\vskip 2mm

{\bf Remarks}. As follows from Eq. (40), the eigenvalues of the
multiplier matrix are given by expression (31). A simple example,
when condition (37) is valid, is the case of a stationary basis
$\{\vp_n\}$, with $\vp_n(t)=\vp_n$ not depending on time.

\vskip 2mm

Comparing Eqs. (3), (13), and (16), we see that
\be
\label{41}
\hat M(t) =\; \ll \hat M(\xi,t)\gg \; .
\ee
Therefore, if $\hat M(\xi,t)$ possesses eigenvectors $\vp_n(t)$, then 
the matrix (41) satisfies the eigenproblem 
\be
\label{42}
\hat M(t) \vp_n(t) = \mu_n(t)\vp_n(t)
\ee
with the same eigenvectors and the eigenvalues
\be
\label{43}
\mu_n(t) = \; \ll \mu_n(\xi,t)\gg \; ,
\ee
which have the property
\be
\label{44}
\mu_n(0) = 1 \; .
\ee

With the spectral norm
$$
||\hat M(t)|| = \sup_n | \mu_n(t)| \; ,
$$
the local stability exponent (15) becomes
\be
\label{45}
\sgm(t) = \ln\sup_n |\ll \mu_n(\xi,t)\gg | \; .
\ee
In this way, the problem of analyzing the stability of a stochastic
dynamical system is connected with finding the eigenvalues of the 
stochastic multiplier matrix.

\section{Concept of Quasi-Isolated Systems}

As is discussed in the Introduction, no real physical system can be 
completely isolated from its surrounding. The latter can be modelled 
by stochastic perturbations of the system dynamics. To stress that the
amplitude of the stochastic perturbation is small, it is convenient
to include explicitly a small factor $\al$ in front of the stochastic
field $\xi(t)$. So, instead of Eq. (1), we shall write
\be
\label{46}
\frac{dy}{dt} =  v(y,\al\xi,t) \; .
\ee
The factor $\al=\al_1+i\al_2$ is assumed to be complex, with its 
real part $\al_1\equiv{\rm Re}\; \al$ and imaginary part 
$\al_2\equiv{\rm Im}\; \al$. The complex value of the factor $\al$ 
makes it possible to simulate random fluctuations of different 
physical quantities, such as energy and attenuation or density and 
phase. If $\al\equiv 0$, there are no stochastic perturbations, and 
one returns to a deterministic dynamical system. When stochastic 
fields are switched on by means of $\al\not\equiv 0$, we have a 
stochastic dynamical system, whose local stability is characterized 
by the stability exponent (45) that takes the form
\be
\label{47}
\sgm(\al,t) \equiv \ln\sup_n |\ll \mu_n(\al\xi,t)\gg | \; ,
\ee
where the dependence on the switching factor $\al$ is explicitly
shown.

The stability exponent (47) describes the stability of a stochastic 
dynamical system with respect to the infinitesimal variation of initial
conditions. For correctly defining the notion of a quasi-isolated 
system, it is also necessary to consider the stability with respect
to infinitesimal stochastic perturbations. This implies that, 
{\it after} analyzing the stability of the stochastic system by means 
of the stability exponent (47), we should set $\al\ra 0$. Since $\al$
is complex-valued, the limit $\al\ra 0$ means that both its real and
imaginary parts tend to zero: $\al_1\ra 0$ and $\al_2\ra 0$. Among
all admissible ways of tending to zero for $\al\ra 0$, it is necessary
to chose that one providing the maximal value for the exponent (47),
in agreement with its definition as characterizing the {\it largest}
deviation of the trajectory. The so defined limit $\al\ra 0$ will be
denoted as
\be
\label{48}
\lim_{\al\ra 0} \sgm(\al,t) \equiv \sup_\al \; \lim_{|\al|\ra 0}
\sgm(\al,t) \; .
\ee

In the stability analysis with the help of the stability exponent (47), 
an important part is the consideration of the asymptotic stability, 
when $t\ra\infty$. This limit may, in general, not commute with the
limit $\al\ra 0$. Therefore, an important step is to study the
commutativity of these limits, characterized by the commutator
$$
[\lim_{\al\ra 0}, \; \lim_{t\ra\infty}] \equiv \lim_{\al\ra 0}
\lim_{t\ra\infty} - \lim_{t\ra\infty}\lim_{\al\ra 0} \; .
$$
The content of this section can be summarized by formulating the
following definitions.

\vskip 2mm

{\bf Definition 1}. A physical system is called {\it quasi-isolated}
if its evolution is described by the stochastic dynamical system (46) 
with infinitesimally small stochastic perturbations.

\vskip 2mm

{\bf Definition 2}. A quasi-isolated system is {\it stochastically
stable} when
\be
\label{49}
[\lim_{\al\ra 0},\; \lim_{t\ra\infty}]\sgm(\al,t) =   0 \; .
\ee

\vskip 2mm

{\bf Definition 3}. A quasi-isolated system is {\it stochastically
unstable} if
\be
\label{50}
[\lim_{\al\ra 0},\; \lim_{t\ra\infty}]\sgm(\al,t) \neq 0 \; .
\ee

Note that the inclusion of stochastic fields in the evolution 
equations can be realized in different ways. Hence, in principle, 
one could consider the stochastic stability with respect to each of 
particular ways. A quasi-isolated system may turn to be stochastically 
stable with respect to some of perturbations but unstable with respect 
to others. However, the kind of action of random environment on a 
quasi-isolated system is, by assumption, unpredictable. Therefore,
it is not sufficient to limit ourselves by only some ways of including
stochastic perturbations, which would result in the analysis of 
{\it partial} stochastic stability. But, in order to make conclusion 
on the {\it general} stochastic stability of a quasi-isolated system,
one must analyze all qualitatively different admissible ways of
including stochastic terms in the evolution equations. Fortunately,
there are just two main qualitatively different types of random noise,
multiplicative and additive.

In the following sections, the preceding ideas will be illustrated  
by concrete examples. Since the perturbing influence of surrounding 
may be caused by many independent random sources, their action, 
according to the central limit theorem, can be modelled by the 
Gaussian white noise [18]. For the convenience of the reader, 
the basic properties of this noise, which will be repeatedly used 
throughout the paper, are listed in short in the Appendix.

\section{Importance of Multiplicative Noise}

One may notice that additive noise cannot lead to stochastic 
instability. Really, let the velocity field in Eq. (1) be a sum 
$v(y,\xi,t)=v_1(y,t)+v_2(\xi,t)$ of two terms, the first of which 
does not depend on the stochastic field $\xi(t)$, while the second 
does not include the dynamic state $y$. Then the Jacobian matrix (19)
is defined only through the variation of $v_1$ and does not depend on 
$v_2$. Therefore the solution of Eq. (20) for the multiplier matrix
also is independent from $v_2$, which means that $v_2$ does not
influence the properties of the multiplier matrix, hence, does not
change the type of stability.

But the multiplicative noise can strongly influence the stability
property. To illustrate this, let us consider the evolution equation
(46) with the velocity field
$$
v(y,\al\xi,t) = f(t) +\al\xi(t) y(\al\xi,t) \; ,
$$
where $f(t)$ is a given function and $\xi(t)$ is a Gaussian white-noise
variable with the properties described in the Appendix. The equation 
(46),
\be
\label{51}
\frac{dy}{dt} = f(t) +\al\xi(t) y \; ,
\ee
determines the evolution of a one-dimensional dynamical system. In this 
case, the Jacobian matrix (19) reduces to the function
$$
J(\al\xi,t) = \al\xi(t) \; .
$$
According to Eq. (31), this gives the multiplier
\be
\label{52}
\mu(\al\xi,t) = \exp\left\{ \al\int_0^t \xi(t')\; dt'\right\} \; .
\ee
The same form (52) could be obtained from the direct variation of the
solution
$$
y(\al\xi,t) = y(0)\exp\left\{ \al \int_0^t \xi(t')\; dt'\right \} 
+ \int_0^t f(t') \exp\left\{ \al \int_{t'}^t \xi(t'')\; dt''
\right\} \; dt' \; .
$$
For the stability index (47), we find
\be
\label{53}
\sgm(\al,t) = (\al_1^2 - \al_2^2) \gm t \; ,
\ee
where $\al_1\equiv {\rm Re}\;\al$ and $\al_2\equiv{\rm Im}\;\al$. 
Keeping in mind the definition (48), according to which the stability 
index is to be maximized with respect to $\al_1$ and $\al_2$, under 
the given modulus $|\al|^2=\al_1^2 +\al_2^2$, we see that 
$\sup_\al(\al_1^2-\al_2^2)$ equals $|\al|^2=\al_1^2$. Therefore the 
index (53) can be written as
\be
\label{54}
\sgm(\al,t) = |\al|^2 \gm t \; .
\ee
From here it follows that the limits
\be
\label{55}
\lim_{t\ra\infty}\;\lim_{\al\ra 0} \sgm(\al,t) = 0 \; , \qquad
\lim_{\al\ra 0}\;\lim_{t\ra\infty} \sgm(\al,t)  = \infty \; 
\ee
do not commute with each other. This implies that the quasi-isolated 
system, whose evolution is given by Eq. (51), is stochastically
unstable.

\section{Oscillator in Stochastic Background}

Many physical processes are presented by oscillatory motion. It is,
therefore, illustrative to consider a quasi-isolated system described
by a harmonic oscillator subject to the action of a weak external
noise. Let the evolution equation (46) have the form
\be
\label{56}
\frac{dy}{dt} = i\om y + \al\xi(t) y \; ,
\ee
where the oscillator frequency $\om$ is real. Here the real part of
$\al$ corresponds to the noisy attenuation-generation process and the
imaginary part of $\al$ described the noise of frequency.

For this one-dimensional case, the Jacobian matrix (19) is the function
$$
J(\al\xi,t) = i\om + \al\; \xi(t) \; .
$$
In view of Eq. (31), the multiplier is
\be
\label{57}
\mu(\al\xi,t) = \exp\left\{ i\om t +\al \int_0^t \xi(t')\; dt'
\right \} \; .
\ee
The same expression (57) also follows from the variation of the 
solution
$$
y(\al\xi,t) = y(0)\exp\left\{ i\om t + \al \int_0^t \xi(t')\; dt'
\right\} \; .
$$
The stability index (47) is
\be
\label{58}
\sgm(\al,t) = \left ( \al_1^2 - \al_2^2\right ) \gm t \; ,
\ee
where the properties of the white noise from the Appendix are used.

If the influence of the random noise is removed before the temporal
limit, that is, $\al\ra 0$, then, for any choice of $\al_1$ and
$\al_2$, we have
\be
\label{59}
\lim_{t\ra\infty}\; \lim_{\al\ra 0} \sgm(\al,t) = 0 \; ,
\ee
which corresponds to the neutral motion. However, the situation is
different if the limit $t\ra\infty$ is taken first. Then, maximizing
the factor (58), in agreement with definition (48), as is explained in 
the previous section, we get the form (54). As s result,
\be
\label{60}
\lim_{\al\ra 0}\; \lim_{t\ra\infty} \sgm(\al,t) = \infty \; .
\ee
The noncommutativity of the limits (59) and (60) shows that the
oscillatory motion is stochastically unstable.

This means that for a finite time, such that $|\al|^2\gm t\ll 1$, the 
system with an oscillatory evolution can approximately be treated as
isolated. But there always exists such a weak random noise that
makes the system unstable for sufficiently long times.

\section{Stochastic Diffusion Equation}

Consider the diffusion equation
\be
\label{61}
\frac{\prt y}{\prt t}  =\left [ D + \al\xi(t)\right ] 
\frac{\prt^2 y}{\prt x^2} \; ,
\ee
in which the diffusion constant $D>0$ is subject to weak random
fluctuations. For any given finite interval, the spatial variable
$x$ can always be scaled so that to be defined on the unity interval. 
Thus, we assume that $x\in[0,1]$. Equation (61) is complimented by 
the initial condition
\be
\label{62}
y(x,\al\xi,0) = y(x,0) \; ,
\ee
with a given function $y(x,0)$, and by the boundary conditions
\be
\label{63}
y(0,\al\xi,t) = b_0 \; , \qquad y(1,\al\xi,t) = b_1 \; ,
\ee
where $b_0$ and $b_1$ are constant.

	For Eq. (61), the Jacobian matrix (19) is 
\be
\label{64}
J(x,x',\al\xi) =\left [ D + \al\xi(t) \right ] 
\frac{\prt^2}{\prt x^2}\; \dlt(x-x') \; .
\ee
The boundary conditions (63) lead, according to Eqs. (22) and (23),
to the boundary conditions
\be
\label{65}
M(0,x',\al\xi,t) = M(1,x',\al\xi,t) = 0 
\ee
for the multiplier matrix.

Solving the eigenproblem
\be
\label{66}
\int_0^1 J(x,x',\al\xi)\vp_n(x')\; dx'=  J_n(\al\xi)\vp_n(x)
\ee
for the Jacobian matrix (64), with the boundary conditions
\be
\label{67}
\vp_n(0) = \vp_n(1)  = 0 \; ,
\ee
we find the eigenvalues
\be
\label{68}
J_n(\al\xi) = - \left [ D + \al\xi(t)\right ] k_n^2
\ee
and the eigenfunctions
\be
\label{69}
\vp_n(x) = \sqrt{2}\; \sin k_n x \; ,
\ee
where
\be
\label{70}
k_n \equiv \pi n \qquad (n=1,2,\ldots,N\ra\infty) \; .
\ee

The eigenvectors $\vp_n=[\vp_n(x)]$, being the columns with the 
elements (69), are stationary. Hence, they satisfy condition (37).
Then, by theorem 3, the multiplier matrix possesses the same 
eigenvectors $\vp_n$, with the eigenvalues (31), where $J_{nn}=J_n$.
Taking account of Eq. (68) yields
\be
\label{71}
\mu_n(\al\xi,t) =\exp\left\{ - Dk_n^2 t - \al k_n^2
\int_0^t \xi(t')\; dt'\right \} \; .
\ee

Note that the solution to Eq. (61) reads 
$$
y(x,\al\xi,t) = \sum_{n=1}^\infty c_n\mu_n(\al\xi,t)\vp_n(x) +
f(x) \; ,
$$
where
$$
c_n =\int_0^1 [ y(x,0) - f(x) ]\vp_n(x)\; dx \; , \qquad
f(x) = b_0 +(b_1 -b_0) x\; .
$$
The form of this solution is that of the expansion (27) in theorem 1,
because of which the multiplier matrix could be found by means of this 
theorem.

Averaging Eq. (71) over the stochastic field (see Appendix), we get
\be
\label{72}
|\ll \mu_n(\al\xi,t) \gg | =\exp\left ( - D k_n^2 t + \al^2 k_n^4
\gm t\right ) \; ,
\ee
where $\al$ is real. Hence, the stability index (47) becomes
\be
\label{73}
\sgm(\al,t) = \sup_n \left ( - Dk_n^2 t +\al^2 k_n^4 \gm t\right ) \; .
\ee
Taking into consideration Eq. (70), this gives
\begin{eqnarray}
\sgm(\al,t) =\left \{ \begin{array}{cc}
-D\pi^2 t & (\al=0) \\
\nonumber
\al^2(\pi N)^4\gm t & (\al\neq 0) \end{array}\right.
\end{eqnarray}
with $N\ra\infty$.

In this way, we have
\be
\label{74}
\lim_{t\ra\infty}\; \lim_{\al\ra 0} \sgm(\al,t) = -\infty \; ,
\ee
which means that in the absence of any stochastic perturbations the
motion would be stable. However, if infinitesimally small stochastic
perturbations are present, then
\be
\label{75}
\lim_{\al\ra 0} \; \lim_{t\ra\infty} \sgm(\al,t) = \infty \; ,
\ee
and the motion is stochastically unstable. This case serves as a 
good example of how even very weak perturbations can render the 
system to become unstable, even if without these perturbations it was 
perfectly stable.

\section{Stochastic Schr\"odinger Equation}

Consider the nonstationary Schr\"odinger equation
\be
\label{76}
\frac{\prt\psi}{\prt t} = \left [ - i H(\br) + \al f(\br,t)\xi(t) 
\right ] \psi \; ,
\ee
in which we set $\hbar\equiv 1, \; \psi=\psi(\br,\al\xi,t)$ is a wave
function, $H(\br)$ is a Hamiltonian, $\al$ is real, $f(\br,t)$ is
a given real function, and $\xi(t)$ is the white noise. With the 
velocity field defined by the right-hand side of Eq. (76), the Jacobian
matrix (19) becomes
\be
\label{77}
J(\br,\br',\al\xi,t) = \left [ - i H(\br) + \al f(\br,t)\xi(t)
\right ] \dlt(\br-\br') \; .
\ee
The eigenproblem for the matrix $\hat J(\al\xi,t)$, whose elements 
are given by Eq. (77), reads
\be
\label{78}
\hat J(\al\xi,t)\psi_n = J_n(\al\xi,t) \psi_n \; .
\ee
Keeping in mind that $\al$ is small, the eigenproblem (78) can be 
solved by means of perturbation theory. In the zero approximation, 
the eigenvector $\psi_n=[\psi_n(\br)]$ is a column with respect to 
the spatial variable $\br$, with $\psi_n(\br)$ given by the stationary
Schr\"odinger equation
$$
H(\br)\psi_n(\br) = E_n\psi_n(\br) \; ,
$$ 
so that the zero-order eigenvalue of the Jacobian matrix is
$$
J_n^{(0)}(\al\xi,t) = - iE_n \; .
$$
The first-order approximation for the eigenvalue of the Jacobian 
matrix is given by
\be
\label{79}
J_n(\al\xi,t) = \psi_n^+ \hat J(\al\xi,t) \psi_n \; ,
\ee
whish yields
\be
\label{80}
J_n(\al\xi,t) = - iE_n + \al f_n(t) \xi(t) \; ,
\ee
where
$$
f_n(t) \equiv \int \psi_n^*(\br) f(\br,t) \psi_n(\br)\; d\br \; .
$$
Note that if $f(\br,t)=f(t)$ does not depend on the spatial variable
$\br$, then the form (80) with $f_n(t)=f(t)$ is an exact eigenvalue
of the matrix $\hat J(\al\xi,t)$. The multi-index $n$, labelling the
eigenvalues, can be discrete as well as continuous.

For the stationary eigenvectors $\psi_n$ of the Jacobian matrix, the
multiplier matrix, by theorem 3, possesses the same eigenvectors and 
its eigenvalues are
\be
\label{81}
\mu_n(\al\xi,t) =\exp\left\{ - i E_n t + \al
\int_0^t f_n(t')\xi(t') \; dt'\right \} \; .
\ee
From here, the stability index (47) is
\be
\label{82}
\sgm(\al,t) = \al^2 \gm \int_0^t f_n^2(t') \; dt'\; .
\ee
The function $f(\br,t)$ in Eq. (76) can always be chosen so that
to satisfy the inequality
\be
\label{83}
\lim_{t\ra\infty} \; \frac{1}{t}\; \int_0^t f_n^2(t') \; dt' > 0 \; .
\ee
Switching off stochastic fields results in the neutral motion, for 
which
\be
\label{84}
\lim_{t\ra\infty} \; \lim_{\al\ra 0} \sgm(\al,t) = 0 \; .
\ee
But for infinitesimally weak stochastic perturbations, the motion 
becomes unstable, with
\be
\label{85}
\lim_{\al\ra 0} \; \lim_{t\ra\infty} \sgm(\al,t) = \infty \; ,
\ee
where condition (83) is taken into account. In this way, the system 
described by the Schr\"odinger equation is stochastically unstable, 
although for some temporal interval, when $\sgm(\al,t)\ll 1$, it can 
be treated as almost isolated.

\section{Sketch of General Situation}

In the general case, the stochastic field $\xi(t)=[\xi_i(x,t)]$ is
a column composed of the elements $\xi_i(x,t)$ depending on space as
well as on time. This field has to enter the evolution equations as a
multiplicative noise. To consider a quasi-isolated system, the
stochastic term is included with the factor $\al$, which is assumed 
to be infinitesimally small. For $\al\ll 1$, the Jacobian matrix (19)
can be calculated by perturbation theory, which yields an expression 
of the form
$$
\hat J(\al\xi,t) \simeq \hat J(0,t) +\al \hat J'(\xi,t) \; .
$$
In the representation of a basis $\{\vp_n(t)\}$ of vectors 
$\vp_n(t)=[\vp_{ni}(x,t)]$, this reads
\be
\label{86}
J_{mn}(\al\xi,t) \simeq J_{mn}(0,t) + \al
\sum_i \int A_{mn}^i (x,t)\xi_i(x,t)\; dx \; .
\ee
If $\vp_n(t)$ are the eigenvectors of the multiplier matrix 
$\hat M(\al\xi,t)$, then its eigenvalues, by theorem 2, are given by 
Eq. (31). Averaging these eigenvalues over stochastic fields, implied
to be Gaussian, gives 
\be
\label{87}
\ll \mu_n(\al\xi,t) \gg \; = \mu_n(0,t) \exp\left\{
\frac{\al^2}{2}\;\kappa_n(t)\right \} \; ,
\ee
where the factor
$$
\mu_n(0,t) =\exp\left\{ \int_0^t J_{nn}(0,t')\; dt'\right \}
$$
is the multiplier of an isolated system, without any random 
perturbations, and
$$
\kappa_n(t) =\sum_{ij} \int dx_1\; dx_2 \; \int_0^t A_{nn}^i(x_1,t_1)
A_{nn}^j(x_2,t_2)\ll \xi_i(x_1,t_1)\xi_j(x_2,t_2)\gg \;
dt_1 \; dt_2
$$
is caused by stochastic perturbations. Then the stability index (47) 
is
\be
\label{88}
\sgm(\al,t) = \sup_n  {\rm Re}\left [ \int_0^t J_{nn}(0,t')\;
dt' + \frac{\al^2}{2}\; \kappa_n(t)\right ] \; ,
\ee
where $\al$ is real.

When $\al\ra 0$, the stability index
\be
\label{89}
\sgm(0,t) =\sup_n\; {\rm Re} \int_0^t J_{nn}(0,t')\; dt'
\ee
is defined by the properties of the system without perturbations. 
The limits $\al\ra 0$ and $t\ra\infty$ do not commute if
\be
\label{90}
\lim_{t\ra\infty}\left |
\frac{\sup_n{\rm Re}\;\kappa_n(t)}{\sgm(0,t)}\right | = \infty \; .
\ee
Then the quasi-isolated system is stochastically unstable. This 
condition is accomplished for the concrete cases considered above. 
It is, of course, impossible to prove that any given quasi-isolated
system is, with probability one, stochastically unstable. However, the
above consideration suggests, with a high level of probability, that 
there always exists such a noise which renders stochastically 
unstable any particular system. This thesis is certainly correct for 
those systems which, in the absence of noise, display neutral motion. 
Then ${\rm Re}\;J_{nn}(0,t)=0$, hence $\sgm(0,t)=0$, and condition (90)
is obviously valid. As is shown in Section 7, condition (90) can be
hold true even for systems that are stable when there is no noise. Let
us also emphasize that if, instead of white noise, we would consider 
infrared noise, then condition (90) would necessarily hold. Really,
for a deterministic system at large time, one usually has 
$\sgm(0,t)\sim t$, while for infrared noise $\kappa_n(t)\sim t^2$. 
This makes condition (90) evidently valid.

Finally, it is important to note that for stochastic dynamical systems
the divergence of averaged trajectories is not necessarily exponential
but may be of algebraic form [4], which implies that the norm of the 
averaged stochastic multiplier matrix has the power-law behaviour
$$
||\ll \hat M(\al\xi,t)\gg || = \al A  \left ( \frac{t}{t_c}
\right )^\bt \qquad (\bt > 0) \; ,
$$
where $A$ is a constant and $t_c$ is the {\it chaotization time} 
defining the crossover between stable and chaotic motion. For $t\ll t_c$,
the motion is stable, while for $t\gg t_c$, it becomes chaotic. The 
arising instability corresponds to {\it weak chaos} since the effective
trajectory divergence is only algebraic but not exponential. In this 
case, the local stability index (47) is
$$
\sgm(\al,t) = \ln \left | \al A\left ( \frac{t}{t_c}
\right )^\bt \right | \; .
$$
From here it follows that the limits
$$
\lim_{t\ra\infty}\; \lim_{\al\ra 0} \sgm(\al,t) = -\infty \; , \qquad 
\lim_{\al\ra 0}\; \lim_{t\ra\infty} \sgm(\al,t) = +\infty 
$$
do not commute with each other. Therefore such a system is also 
stochastically unstable.

\section{Conclusions}

A convenient characteristic for analyzing the stability of dynamical
systems is the {\it local stability index} (5). This can be expressed
through the multiplier matrix $\hat M(t)$ as 
$$
\sgm(t)=\ln||\hat M(t)||\; .
$$
For deterministic (nonstochastic) dynamical systems, there exists 
another representation of the stability index through the Lyapunov or
stability matrix ${\rm Re}\hat J(t)$, where $\hat J(t)$ is the
Jacobian matrix associated with the considered system. The name of the
Lyapunov matrix comes from the fact that its eigenvalues are the local
Lyapunov exponents. Then the stability index, if condition (37) holds,
can be written as
$$
\sgm(t) = \int_0^t ||{\rm Re} \; \hat J(t') ||\; dt'\; .
$$
This presentation, however, is not valid for stochastic dynamical 
systems. For the latter, the stability index is to be calculated by 
means of Eq. (45). The form (47) of the stability index, 
$$
\sgm(\al,t) = \ln || \ll \hat M(\al\xi,t) \gg || \; ,
$$
is a handy representation for studying the influence of weak stochastic
perturbations. The main physical conclusions resulting from the
general approach and particular examples are as follows.

\vskip 2mm

(i) {\it Nonexistence of isolated systems}. The fact that no real 
physical system can be completely isolated, but is always subject to
uncontrollable random perturbations, is more or less generally accepted
[5--11]. The point that the concept of an isolated system is logically 
self-contradictory has also been emphasized [9,12,13]. What is 
principally new in the present paper is the demonstration that isolated
systems are stochastically unstable with respect to infinitesimally
weak random perturbations. A given physical system can be considered as 
almost isolated, or quasi-isolated, during a finite time interval, but
it cannot be treated as such for ever. Sooner or later, a quasi-isolated
system looses its stability. There are no eternally stable systems in 
nature.

\vskip 2mm

(ii) {\it Absence of absolute equilibrium}. In the theory of dynamical
systems, solutions are termed equilibrium if they are either constant
in time or periodic or quasiperiodic. However, for a quasi-isolated
system, no one of these solutions can be absolutely stable for 
infinitely long time. On a finite temporal interval, a solution can 
correspond to a stable equilibrium, but with increasing time, some kind
of nonequilibrium behaviour will certainly appear. For instance, big
fluctuations, driving the system far from equilibrium, may arise
[33,34]. Since statistical systems are a particular type of real
physical systems, they also have to be considered as quasi-isolated.
The absence of absolute equilibrium for a statistical system implies
that large nonequilibrium fluctuations of mesoscopic scale spontaneously
appear in the system, being randomly distributed in space and in time
[35]. If evolution equations do possess an attractor, this has to be a
chaotic attractor.

\vskip 2mm

(iii) {\it Irreversibility of time arrow}. As far as completely 
isolated systems do not exist, but there are only quasi-isolated
systems, the dynamics of such a system, because of the action of 
random perturbations, can never be reversed so that to exactly return 
to a particular dynamical state. Since quasi-isolated systems are 
stochastically unstable, any trajectory after sufficiently long time
will deviate arbitrarily far from the initial point. All that means the
irreversibility of time.

\vskip 5mm

{\bf Acknowledgement}

\vskip 2mm

This work has been supported by the research grants from the S\~ao 
Paulo State Research Foundation, Brazil and Bogolubov-Infeld 
International Program, Poland.

\newpage

{\large{\bf Appendix}}.

\vskip 5mm
Here several formulas, related to the Gaussian white noise, are 
presented, which have been repeatedly used throughout the paper.
The stochastic variable $\xi(t)$, corresponding to this noise, centered
at zero, has the properties
$$
\ll \xi(t)\gg\; = 0 \; , \qquad \ll \xi(t)\xi(t')\gg\; = 2\gm
\dlt(t-t') \; ,
$$
$$
\ll \xi(t_1)\xi(t_2)\ldots\xi(t_{2n+1})\gg\; = 0 \; , 
$$
$$
\ll \xi(t_1)\xi(t_2)\ldots\xi(t_{2n})\gg\; = (2\gm)^n
\sum_{sym}^{(2n-1)!!} \dlt(t_1-t_2)\dlt(t_3-t_4) \ldots
\dlt(t_{2n-1} -t_{2n}) \; ,
$$
where $\sum_{sym}$ implies the symmetrized sum and $(2n-1)!! = (2n)!/2^n n! 
= 1\cdot 3\cdot 5\cdot \cdot\cdot (2n-1)$. As an example, a symmetrized
sum of $f_{ij}$ for $n=2$ means $f_{12}f_{34}+f_{13}f_{24}+f_{14}f_{23}$.
The integration of $\xi(t)$ over time gives the Wiener variable
$$
w(t) \equiv \int_0^t \xi(t')\; dt'\; .
$$
For the latter, one has
$$
\ll \int_{t_1}^{t_2} w(t) \; dw(t) \gg \; = \gm(t_2 - t_1) \; ,
\qquad \ll w^{2n+1}(t)\gg \; = 0\; , \qquad
\ll w^{2n}(t)\gg \; = \frac{(2n)!}{n!}(\gm t)^n \; .
$$
In general, any Gaussian variable $G(t)$ satisfies the equality
$$
\ll \exp G(t)\gg \; = \exp\left \{ \frac{1}{2}\;
\ll G^2(t)\gg \right \} \; .
$$
For instance,
$$
\ll \exp\{\al w(t)\}\gg \; = \exp(\al^2\gm t) \; .
$$
These formulas are sufficient to understand all calculations, related
to the averaging over the white noise, which have been made in the 
paper.


\newpage

\begin{references}
\bibitem{1}
V.V. Nemytskii and V.V. Stepanov, {\it Qualitative Theory of 
Differential Equations} (Princeton University, Princeton, 1960).

\bibitem{2}
J. Guckenheimer and P.J. Holmes, {\it Nonlinear Oscillations, Dynamical
Systems, and Bifurcations of Vector Fields} (Springer, New York, 1986).

\bibitem{3}
A.J. Lichtenberg and M.A. Liberman, {\it Regular and Chaotic Dynamics}
(Springer, New York, 1992).

\bibitem{4}
X. Mao, {\it Stability of Stochastic Differential Equations with
Respect to Semimartingales} (Longman, Harlow, 1991).

\bibitem{5}
D. ter Haar, {\it Elements of Statistical Mechanics} (Reinehart, New 
York, 1954).

\bibitem{6}
L.D. Landau and E.M. Lifshits, {\it Statistical Mechanics} (Pergamon,
Oxford, 1958).

\bibitem{7}
J.E. Mayer and M.G. Mayer, {\it Statistical Mechanics} (Wiley, New
York, 1977).

\bibitem{8}
O. Penrose, Rep. Prog. Phys. {\bf 42}, 1937 (1979).

\bibitem{9}
V.I. Yukalov, {\it Statistical Green's Functions} (Queen's University,
Kingston, 1998).

\bibitem{10}
W.H. Zurek, Prog. Theor. Phys. {\bf 89}, 281 (1993).

\bibitem{11}
W.H. Zurek and J.P. Paz, Phys. Rev. Lett. {\bf 72}, 2508 (1994).

\bibitem{12}
V.I. Yukalov, Mosc. Univ. Phys. Bull. {\bf 25}, 49 (1970).

\bibitem{13}
V.I. Yukalov, Mosc. Univ. Phys. Bull. {\bf 26}, 22 (1971).

\bibitem{14}
J. von Neumann, Mathematical Foundations of Quantum Mechanics 
(Princeton University, Princeton, 1955).

\bibitem{15}
J.L. Lebowitz, Physica A {\bf 263}, 516 (1999).

\bibitem{16}
G.M. Zaslavsky, Phys. Today N8, 39 (1999).

\bibitem{17}
V.I. Yukalov, Physica A {\bf 234}, 725 (1997).

\bibitem{18}
C.W. Gardiner, {\it Handbook of Stochastic Methods} (Springer, Berlin,
1985).

\bibitem{19}
R.L. Stratonovich, {\it Nonlinear Nonequilibrium Thermodynamics}
(Springer, Berlin, 1992).

\bibitem{20}
V.I. Yukalov, Phys. Rev. A {\bf 56}, 5004 (1997).

\bibitem{21}
V.I. Yukalov, Laser Phys. {\bf 7}, 998 (1997).

\bibitem{22}
G.M. Zaslavsky and B.A. Niyazov, Phys. Rep. {\bf 283}, 73 (1997).

\bibitem{23}
D.E. Feldman, Int. J. Mod. Phys. B {\bf 15}, 2945 (2001).

\bibitem{24}
A.J. Coleman and V.I. Yukalov, {\it Reduced Density Matrices} 
(Springer, Berlin, 2000).

\bibitem{25}
H. Fujisaka, Prog. Theor. Phys. {\bf 70}, 1264 (1983).

\bibitem{26}
P. Grassberger, R. Badii, and A. Politi, J. Stat. Phys. {\bf 51}, 
135 (1988).

\bibitem{27}
P.G. Drazin, {\it Nonlinear Systems} (Cambridge University, Cambridge,
1994).

\bibitem{28}
M.B. Priestly, {\it Nonlinear and Nonstationary Time Series Analysis}
(Academic, London, 1988).

\bibitem{29}
V.I. Yukalov, J. Math. Phys. {\bf 32}, 1235 (1991).

\bibitem{30}
V.I. Yukalov, J. Math. Phys. {\bf 33}, 3994 (1992).

\bibitem{31}
V.I. Yukalov and E.P. Yukalova, Physica A {\bf 225}, 336 (1996).

\bibitem{32}
V.I. Yukalov and E.P. Yukalova, Ann. Phys. (N.Y.) {\bf 277}, 219 (1999).

\bibitem{33}
D.G. Luchinsky, P. McKlintock, and M.I. Dykman, Rep. Prog. Phys. 
{\bf 61}, 889 (1998).

\bibitem{34}
B.V. Chirikov and O.V. Zhirov, J. Exp. Theor. Phys. {\bf 120}, 
214 (2001).

\bibitem{35}
V.I. Yukalov, Phys. Rep. {\bf 208}, 395 (1991).

\end{references}
\end{document}